\documentclass[aps,onecolumn,floatfix,superscriptaddress]{revtex4}

\usepackage{amsmath}
\usepackage{amssymb}
\usepackage{times}
\usepackage{braket}
\usepackage{units}
\usepackage[pdftex]{graphicx}
\usepackage{xcolor}

\renewcommand{\vec}[1]{\boldsymbol{#1}}

\DeclareMathOperator{\sign}{sign}

\begin{document}

\title{Signatures of lattice geometry in quantum and topological Hall effect}

\author{B{\"o}rge G{\"o}bel}
\email[]{bgoebel@mpi-halle.mpg.de}
\affiliation{Max-Planck-Institut f\"ur Mikrostrukturphysik, D-06120 Halle (Saale), Germany}

\author{Alexander Mook}
\affiliation{Max-Planck-Institut f\"ur Mikrostrukturphysik, D-06120 Halle (Saale), Germany}

\author{J\"urgen Henk}
\affiliation{Institut f\"ur Physik, Martin-Luther-Universit\"at Halle-Wittenberg, D-06099 Halle (Saale), Germany}

\author{Ingrid Mertig}
\affiliation{Max-Planck-Institut f\"ur Mikrostrukturphysik, D-06120 Halle (Saale), Germany}
\affiliation{Institut f\"ur Physik, Martin-Luther-Universit\"at Halle-Wittenberg, D-06099 Halle (Saale), Germany}

\date{\today}

\begin{abstract}
The topological Hall effect (THE) of electrons in skyrmion crystals is strongly related to the quantum Hall effect (QHE) on lattices. This relation suggests to revisit the QHE because its Hall conductivity can be unconventionally quantized.  It exhibits a jump and changes sign abruptly if the Fermi level crosses a van Hove singularity. In this Paper, we investigate the unconventional QHE features by discussing band structures, Hall conductivities, and topological edge states for square and triangular lattices; their origin are Chern numbers of bands in the skyrmion crystal (THE) or of the corresponding Landau levels (QHE). Striking features in the energy dependence of the Hall conductivities are traced back to the band structure without magnetic field whose properties are dictated by the lattice geometry. Based on these findings, we derive an approximation that allows us to determine the energy dependence of the topological Hall conductivity on \emph{any} twodimensional lattice. The validity of this approximation is proven for the honeycomb lattice. We conclude that skyrmion crystals lend themselves for experiments to validate our findings for the THE and---indirectly---the QHE\@.
\end{abstract}

\maketitle

\section{Introduction}
With the recent ascent of skyrmions~\cite{bogdanov1989thermodynamically,bogdanov1994thermodynamically, rossler2006spontaneous,muhlbauer2009skyrmion,nagaosa2013topological}---particle-like topologically nontrivial field configurations~\cite{skyrme1962unified}---to one of the most auspicious research areas in physics, the transport of electrons in a Hall geometry may become of great interest again. Skyrmions in magnets rely typically on the  Dzyaloshinskii-Moriya interaction~\cite{dzyaloshinsky1958thermodynamic,moriya1960anisotropic} and are detected in non-centrosymmetric B20 materials, e.\,g., in MnSi~\cite{muhlbauer2009skyrmion}. Other mechanisms~\cite{nagaosa2013topological}, e.\,g., frustration \cite{okubo2012multiple}, allow for smaller skyrmions. The skyrmion size is not only relevant for potential applications in storage and spintronics devices~\cite{romming2013writing,zhang2015magnetic,zhang2015magnetic2,hsu2016electric} but also for the magnitude of the skyrmion-induced transport signal; the latter often depends on the skyrmion density
\begin{align}
	n_\mathrm{Sk}(\vec{r}) &= \vec{s}(\vec{r}) \cdot \left( \frac{\partial}{\partial x} \vec{s}(\vec{r}) \times \frac{\partial}{\partial y} \vec{s}(\vec{r}) \right) \label{eq:skyrmiondensity}
\end{align}
[$\vec{s}(\vec{r})$ spin texture of the skyrmion].

The topological Hall effect (THE)~\cite{bruno2004topological,neubauer2009topological,schulz2012emergent,kanazawa2011large, lee2009unusual,li2013robust,hamamoto2015quantized,lado2015quantum,gobel2017THEskyrmion} of electrons in skyrmion crystals---regular arrays of skyrmions---arises from the real-space Berry curvature of the spin texture which produces an emergent magnetic field proportional to $n_\mathrm{Sk}(\vec{r})$. The THE is closely related to the quantum Hall effect (QHE) on lattices~\cite{gobel2017THEskyrmion}. The description of the QHE for free electrons in terms of dispersionless Landau levels (LLs)~\cite{landau1930diamagnetismus} motivated Onsager to formulate a scheme to deduce LLs from any band structure~\cite{onsager1952interpretation}. The experimental discovery~\cite{klitzing1980new} of the QHE showed that this theory is valid in general, except for small deviations associated with the underlying lattice. Hofstadter butterflies calculated for various lattices~\cite{hofstadter1976energy,claro1979magnetic, rammal1985landau, claro1981spectrum, thouless1982quantized} confirmed Onsager's quantization scheme but the LLs did not appear perfectly dispersionless (as is the case for free electrons).

The anomalous quantum Hall conductivity of graphene near half filling~\cite{novoselov2005two} motivated to describe the QHE by means of Chern numbers~\cite{hatsugai2006topological,sheng2006quantum}. It was found that LLs near a van Hove singularity would cause an enormous quantum Hall signal, fully compensating the contributions of all other LLs; such a feature is absent for free electrons and cannot be explained with Onsager's quantization scheme.

\begin{figure*}
  \centering
  \includegraphics[width=0.95\columnwidth]{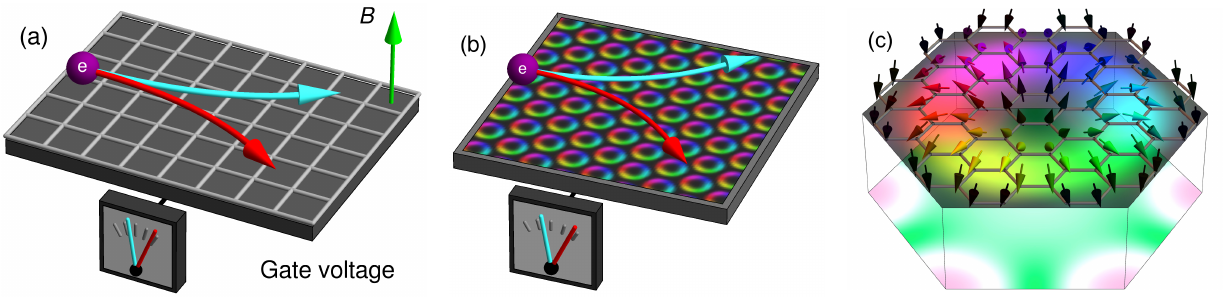}
  \caption{Setups for the quantum and the topological Hall effect. (a) Hall bar ($xy$ plane) with square structural lattice in an external homogeneous magnetic field $\vec{B}$ (green arrow along the $\vec{z}$ direction). The sign of the quantum Hall conductivity can be changed by tuning the gate voltage; this effect is caused by the fermion character of the electrons (electron- versus holelike) that depends on the adjusted Fermi energy. As a result, an electron (sphere) is deflected to the left or to the right (blue and red arrows). (b) Analogous setup for the topological Hall effect. The external magnetic field is replaced by a skyrmion crystal, the latter represented as color-coded circles. (c) Closeup of the magnetic unit cell of a skyrmion on a honeycomb structural lattice. The spin texture is represented in Lorentz-microscopy style. In the top part, the in-plane component of the spins (arrows) is coded by the color scale. The emergent magnetic field is shown in the bottom plane (green: positive, red: negative).}
  \label{fig:systems}
\end{figure*}

Recently we have shown that the THE in a skyrmion crystal can be mapped onto the QHE by homogenization of the emergent field~\cite{gobel2017THEskyrmion}; this correspondence tells that THE and QHE describe essentially the same physics (QHE in Fig.~\ref{fig:systems}a, THE in Fig.~\ref{fig:systems}b). When electrons are strongly coupled to the skyrmion texture, THE experiments could simultaneously verify the validity of the topological theory for the QHE\@. 

Berry curvature and Chern numbers allow for profound understanding of both effects. In this Paper we elaborate on the general nature of the effects and point out the importance of van Hove singularities whose properties are dictated by the structural lattice. In addition, we propose a handy approximation for the energy-dependent Hall conductivity that circumvents calculations of the Berry curvature; its validity is checked for the QHE and the THE on a honeycomb lattice.

This Paper is organized as follows. Theoretical issues are addressed in Section~\ref{sec:theory} in which we recapitulate topological transport (\ref{sec:topological-transport}), as well as the QHE for free electrons (\ref{sec:QHE-free}) and for electrons on a lattice (\ref{sec:QHE-lattice}). In Section~\ref{sec:Results} we present and discuss results for the QHE on a square lattice in detail (\ref{sec:QHEsquare}) and briefly for a triangular lattice (\ref{sec:QHEtriangular}). Subsequently, we turn to the THE in skyrmion crystals (\ref{sec:THE}) and discuss its relation to the QHE\@. Inspired by the close relation of THE and QHE we introduce an approximation for the energy-dependent Hall conductivity of both QHE and THE (\ref{sec:approx}).  We conclude with Section~\ref{sec:conclusion} which is attributed to an experimental verification and motivates further theoretical research.

\section{Theoretical aspects}
\label{sec:theory}

\subsection{Topological contributions to Hall coefficients}
\label{sec:topological-transport}
The twodimensional electronic system in the $xy$ plane is described by a Hamiltonian $H$ in tight-binding formulation (explicit formulations are given below). The Berry connection 
\begin{align*}
  \vec{A}_{n}(\vec{k}) & = \mathrm{i} \braket{u_{n}(\vec{k})|\nabla_{\vec{k}}|u_{n}(\vec{k})} 
\end{align*}
and the Berry curvature
\begin{align*}
  \Omega_{n}^{(z)}(\vec{k}) = \frac{\partial}{\partial k_{x}} A_{n}^{(y)}(\vec{k}) -  \frac{\partial}{\partial k_{y}} A_{n}^{(x)}(\vec{k}) 
\end{align*}
for all bands $n$ are calculated from their eigenvectors $u_{n}(\vec{k})$ with eigenenergies $E_{n}(\vec{k})$. The intrinsic transverse Hall conductivity is given by the Kubo formula~\cite{nagaosa2010anomalous}
\begin{align*}
  \sigma_{xy}(E_\mathrm{F}) & = \frac{e^{2}}{h} \frac{1}{2\pi} \sum_{n} \int_{\mathrm{BZ}} \Omega_{n}^{(z)}(\vec{k}) \, f(E_{n}(\vec{k}) - E_\mathrm{F}) \,\mathrm{d}^{2}k,
\end{align*}
evaluated as a Brillouin zone (BZ) integral; $f(x)$ is the Fermi distribution function. $e$ and $h$ are the electron charge and the Planck constant, respectively. At zero temperature only states below the Fermi energy $E_\mathrm{F}$ contribute to transport: if $E_\mathrm{F}$ is located in the band gap above the $l$-th band, $\sigma_{xy}$ is proportional to the winding number~\cite{Hatsugai1993,Hatsugai1993a}
\begin{align}
  w_{l} & = \sum_{n \le l} C_n,
  \label{eq:winding-number}
\end{align}
in which 
\begin{align*}
  C_{n} & = \frac{1}{2\pi} \int_{\mathrm{BZ}}\Omega_{n}^{(z)}(\vec{k}) \,\mathrm{d}^{2}k
\end{align*}
is the Chern number of the $n$-th band. The winding number tells number and propagation direction of topologically nontrivial edge states within in the $l$-th band gap. More precisely, this bulk-boundary correspondence~\cite{hasan2010colloquium} identifies $w_{l}$  with $n^{\mathrm{R}}_{l}$ edge states with right-handed and $n^{\mathrm{L}}_{l}$  edge states with left-handed chirality,
\begin{align*}
  w_{l} & = n^{\mathrm{R}}_{l} - n^{\mathrm{L}}_{l}.
\end{align*}
These edge states distinguish a topological from a conventional insulator~\cite{hasan2010colloquium,kane2005z,kane2005quantum,bernevig2006quantum}. 

\subsection{Quantum Hall effect for free electrons}
\label{sec:QHE-free}
Free electrons that are confined to the $xy$ plane in a homogeneous magnetic field $\vec{B} = B \vec{e}_{z}$ are described by the Hamiltonian
\begin{align*}
  H & = \frac{1}{2m} (\vec{p} + e \vec{A})^2,
\end{align*}
in which the vector potential $\vec{A}$ defines the magnetic field $\vec{B} = \nabla \times \vec{A}$. A canonical transformation maps this Hamiltonian onto that of a harmonic oscillator, giving dispersionless equidistant LLs with energies~\cite{nolting2009quantum}
\begin{align*}
  E_n & = \hbar \omega_{\mathrm{c}} \left( n + \frac{1}{2} \right), \quad n \ge 0,
\end{align*}
with the cyclotron frequency $\omega_{\mathrm{c}} = e B / m$. A constant-energy cut of the free-electron parabola at $E_n$ encloses the area (in reciprocal space)
\begin{align}
\zeta_n &= \zeta_0 \left( n + \frac{1}{2} \right), \quad \zeta_0 = \frac{B}{\Phi_0}, \quad \Phi_0 = \frac{h}{e}.
\label{eq:numstates}
\end{align}
Therefore, the `number of states' of each LL is identical. The constant Berry curvature
\begin{align*}
  \Omega^{(z)}_n(\vec{k}) & = \Omega_0,
\end{align*}
of a LL (e.\,g., calculated in Landau gauge $\vec{A} = B y \vec{e}_{x}$) yields its Chern number $C_n = -1$. This tells that the number of topological nontrivial edge states  in adjacent band gaps differs by $\pm 1$. The larger $B$, the smaller is the number of edge states below a fixed Fermi level and the smaller is the Hall conductivity, because the energy difference of two adjacent LLs is proportional to $B$.

In Onsager's quantization scheme~\cite{onsager1952interpretation} the above result for the free electron parabola is carried over to any zero-field band structure (calculated for $B = 0$). A Landau level is formed if the enclosed area in reciprocal space fulfills relation \eqref{eq:numstates}. Hence, each LL exhibits the same occupation, as for free electrons.

\subsection{Quantum Hall effect on a lattice}
\label{sec:QHE-lattice}
For electrons on a lattice, the sum over all Chern numbers $C_{n}$ has to be zero. Therefore, lattice properties introduce phenomena that are missing for Landau levels stemming from free electrons.

The electronic structure for a twodimensional lattice is described by the tight-binding Hamiltonian
\begin{align} 
  H & = \sum_{ij} t_{ij} \, c_{i}^\dagger \, c_{j}
  \label{eq:ham_qhe} 
\end{align}
with nearest-neighbor hopping strengths $t_{ij}$ ($i$ and $j$ site indices); $c_{i}^\dagger$ and $c_{i}$ are creation and annihilation operators, respectively. The hopping strengths
\begin{align}
  t_{ij} & = t \,\mathrm{e}^{- \mathrm{i} \varphi_{ij}}, \quad \varphi_{ij} = \frac{e}{\hbar} \int_{\vec{r}_{i} \to \vec{r}_{j}} \vec{A}(\vec{r}) \cdot \mathrm{d}\vec{r},
 \label{eq:hoppingphase}
\end{align}
depend on the vector potential $\vec{A}(\vec{r})$. The integration is along the line that connects site $i$ with site $j$; $t$ is the hopping strength of the zero-field Hamiltonian.

The phases $\varphi_{ij}$ are not gauge-invariant.  The physically relevant quantity is the magnetic flux through the plaquettes of the lattice. Since the flux is proportional to the sum of the `encircling' $\varphi_{ij}$, the phases have to be compatible with the periodicity of the lattice. This imposes specific values on the magnetic field $B$, so that commensurability
\begin{align}
 t_{ij} = t_{(i+n)(j+n)}
 \label{eq:restriction}
\end{align}
is valid for a lattice with $n$ atoms in its unit cell.

\section{Results and discussion}
\label{sec:Results}
In what follows we investigate and explain how lattice effects manifest themselves in the QHE\@. We start with the instructive square lattice (Section~\ref{sec:QHEsquare}) and turn then to the triangular lattice (Section~\ref{sec:QHEtriangular}) in which an unconventionally quantized Hall conductivity shows up.

After revisiting the THE on a triangular lattice (Section~\ref{sec:THE}) we formulate the approximation for the Hall conductivity of both THE and QHE and check its validity for the honeycomb lattice (Section~\ref{sec:approx}).

\subsection{Quantum Hall effect on a square lattice}
\label{sec:QHEsquare}
The square lattice with lattice constant $a$ is defined by its lattice vectors $\vec{a}_1 = a \vec{e}_{x}$ and $\vec{a}_2 = a \vec{e}_{y}$. In Landau gauge, $\vec{A} = B y \vec{e}_{x}$, the QHE Hamiltonian for this lattice takes the matrix form
\begin{align*}
  H & = t
 \begin{pmatrix}
  h_1& \mathrm{e}^{\mathrm{i} a k_y}	& 0 & \dots	 & 0 & \mathrm{e}^{-\mathrm{i} a k_y} \\
  \mathrm{e}^{-\mathrm{i}  a k_y}	& h_2 & e^{\mathrm{i} a k_y}	& \dots  & 0 &  0  \\
  0 & \mathrm{e}^{-\mathrm{i} a k _y}	& h_3	& \dots & 0 & 0   \\
  \vdots	& \vdots & \vdots 	& \ddots & \vdots& \vdots \\
  0 & 0 &0	& \dots & h_{q-1} & \mathrm{e}^{\mathrm{i} a k_y} \\
  \mathrm{e}^{\mathrm{i} a k_y} & 0 & 0	& \dots & \mathrm{e}^{-\mathrm{i} a  k_y} & h_q
  \end{pmatrix},
\end{align*}
with
\begin{align*}
  h_j & = 2 \cos\left( a k_x + 2 \pi \frac{p}{q} j \right). 
\end{align*}
The coprime integers $p$ and $q$ define the strength $B$ of the magnetic field: $p / q = \Phi / \Phi_0$ with $\Phi = B a^2$. For the most part of this Paper we set $p = 1$ because we aim at relating the QHE to the topological Hall effect in skyrmion crystals; for the latter, $p = 1$ corresponds to the topological charge of a skyrmion (Section~\ref{sec:THE}; the case $p > 1$ is briefly discussed in Section~\ref{eq:quantization-p1}).

The restriction of the $t_{ij}$ [eq.~\eqref{eq:restriction}] compels to use a rectangular unit cell with  lattice vectors $\vec{b}_1 = a \vec{e}_{x}$ and $\vec{b}_2 = a q \vec{e}_{y}$. Hence, the magnetic Brillouin zone covers $1 / q$-th of the structural Brillouin zone.

\subsubsection{Landau levels and Hall conductivity}
\begin{figure}
  \centering
  \includegraphics[width=0.45\columnwidth]{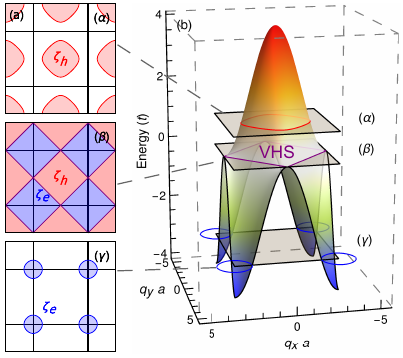}
  \caption{Band structure of a square lattice for $B = 0$. The band structure, depicted in (b), is cut at constant energies $\alpha$, $\beta$, and $\gamma$ (close ups in panel~a). At $\alpha$, there is one closed hole pocket (marked red), whereas for $\gamma$ there is one electron pocket (blue). At the van Hove singularity $\beta$, the band structure exhibits a Lifshitz transition and the fermion character (electron- \textit{versus} holelike) changes. Energies in units of the hopping strength $t$; $a$ lattice constant.}
  \label{fig:QHE_quad_cuts}
\end{figure}

The band structure for $B = 0$ (i.\,e., the zero-field band structure, depicted within the structural Brillouin zone in Fig.~\ref{fig:QHE_quad_cuts}b) has a maximum at $E = +4 \, t$, a minimum at $E = -4 \, t$, and two energetically degenerate VHSs at $E_\mathrm{VHS} = 0$ (for $t>0$); the latter appear as one pole in the density of state and are referred to as `the VHS' in the following.

\begin{figure}
  \centering
  \includegraphics[width = 0.45\columnwidth]{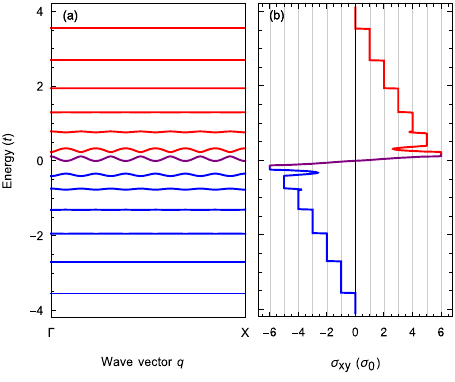}
  \caption{Quantum Hall effect on a square lattice for $p/q = 1/13$. (a) Landau levels. The fermion character is indicated by color: electronlike blue, holelike red (see also Fig.~\ref{fig:QHE_quad_cuts}a). (b) Hall conductivity $\sigma_{xy}$. The conductivity decreases in steps of $-\sigma_0$ at every LL\@. An exception is the  LL at the van Hove singularity (purple, at $E = 0$) that carries a Chern number of $q - 1 = +12$; as a consequence, $\sigma_{xy}$ increases abruptly. Energies in units of the hopping strength $t$; $\sigma_0 = e^{2} / h$.}
  \label{fig:QHE_quad_results}
\end{figure}

For $B > 0$, the emerging Landau levels  are symmetrically distributed about the VHS, which implies that for odd $q$ one LL shows up exactly at the VHS ($q = 13$ in Fig.~\ref{fig:QHE_quad_results}a). On top of this, the LLs exhibit $q$ oscillations. The amplitudes of these oscillations are largest for LLs close to the VHS; on the contrary, LLs close to the band edges appear practically dispersionless. The positions (in reciprocal space) of maxima and minima of every second band coincide. 

The energy-resolved Hall conductivity $\sigma_{xy}$ is zero at energies below the band bottom $E = -4 \, t$ of the zero-field band structure (Fig.~\ref{fig:QHE_quad_results}b). With increasing energy, $\sigma_{xy}$  decreases in steps of $\sigma_0 \equiv e^2 / h$ at each LL, which is readily explained by their Chern numbers of $-1$. These steps comply with LLs of free electrons (Section~\ref{sec:theory}) and are abrupt because the associated LLs are practically dispersionless.

The sizable oscillation amplitudes of the LLs near the VHS at $E_\mathrm{VHS} = 0$ manifest themselves as modulations in $\sigma_{xy}$; in other words, the jumps are not abrupt. This is explained by the Berry curvature which is inhomogeneously distributed within the Brillouin zone, in contrast to the Berry curvature of free-electron LLs. Nevertheless, the Chern numbers equal $-1$.

For odd $q$, the LL closest to the VHS has a Chern number of $q - 1$: this causes a sizable jump and a change of sign in $\sigma_{xy}$. For even $q$, the two LLs closest to the VHS touch each other and carry a joint Chern number of $q - 2$. At even larger energies, $\sigma_{xy}$ decreases and reaches zero at the top of the band structure ($E = + 4 \, t$). 

The overall shape of the energy-resolved conductivity is antisymmetric, which reflects the symmetric shape of the zero-field band structure. Briefly summarizing at this point, LLs and Hall conductivity show features that are clearly attributed to lattice properties.

\subsubsection{Fermion character and Berry curvature}
To elaborate on the above findings we assume that the LLs can be separated into free-electron and lattice-influenced ones. Landau levels of the first type are almost dispersionless, possess an almost constant Berry curvature $\Omega_0$, and have Chern numbers of $-1$. The second type shows up close to $E_\mathrm{VHS}$, with oscillations in both energy and Berry curvature; the positions (in reciprocal space) of their extrema coincide with those of their Berry curvature.

We now discuss the Berry curvature distributions in detail. For this purpose we determine the fermion character of the electrons at constant-energy cuts of the zero-field band structure.

At low energies (cut $\gamma$ in Fig.~\ref{fig:QHE_quad_cuts}) the dispersion is almost parabolic and the circular Fermi line encloses occupied states. This electron pocket has positive curvature and is associated with a positive effective mass $m^\star$. With increasing energy, the dispersion deviates more and more from that of free electrons, the constant energy contours become warped  but the Fermi lines remain electronlike.

\begin{figure*}
  \centering
  \includegraphics[width=1\columnwidth]{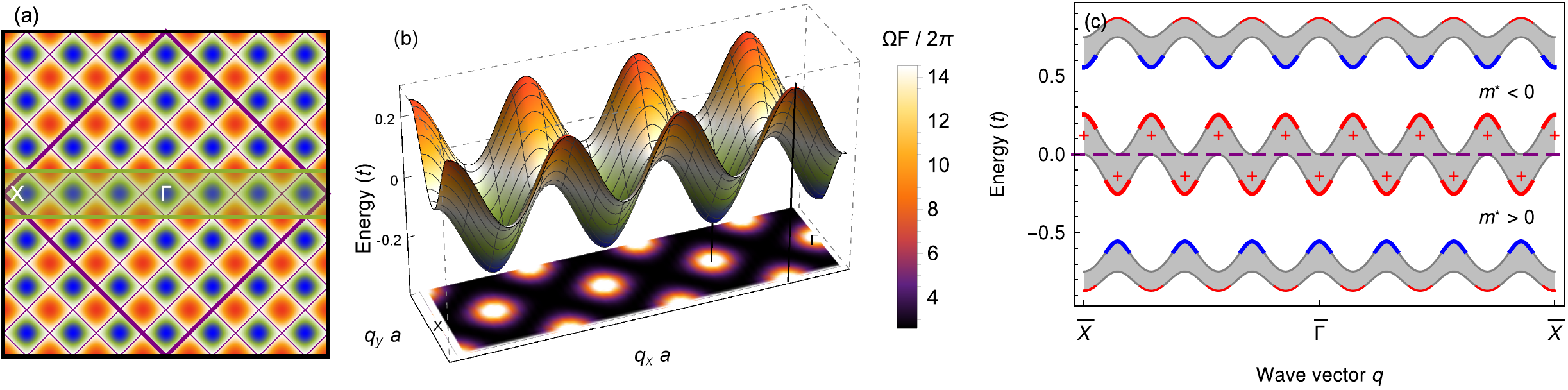}
  \caption{Landau level at the van Hove singularity of a square lattice with $p/q = 1/7$. (a) Lattice-induced modulations of the dispersion relation in the Brillouin zone (BZ) of the square lattice. The magnetic BZ (green) covers $1/q$-th of the original BZ and contains $q$ modulation periods. (b) Dispersion relation (three-dimensional representation) and Berry curvature  $\Omega^{(z)} - \Omega_0$ (contour plot) in the magnetic BZ, normalized to the area of the Brillouin zone $F$ (so that the average gives the Chern number of the band). (c) Schematic presentation of the Berry curvature of the three Landau levels close to the van Hove singularity (dashed purple line). Gray bands are projections onto the one-dimensional cut through the magnetic BZ\@. Extrema of the Berry curvature appear at avoided crossings; its sign is distinguished by color (red: positive, blue: negative).  Energies in units of the hopping strength $t$.}
  \label{fig:VHS_square}
\end{figure*}

At $E_\mathrm{VHS}$ (cut $\beta$) the Fermi line is a square; its vanishing curvature implies an infinite effective mass. Hence, the Lorentz force of an external magnetic field leaves the electronic states unaffected, which explains why the LLs close to $E_\mathrm{VHS}$ show oscillations that resemble the zero-field band structure (cf.\ Figs.~\ref{fig:QHE_quad_cuts}b and~\ref{fig:VHS_square}a). 

At higher energies, the Fermi line becomes circular again but holelike and with negative curvature. In other words, the band structure exhibits a Lifshitz transition~\cite{lifshitz1960anomalies} at $E_\mathrm{VHS}$, which is accompanied by a change of the fermion character: from electronlike below the VHS (with a positive effective mass $m^\star$) to holelike above the VHS (with negative $m^\star$).

The Berry curvature of LLs at the band bottom and at the top of the bands is almost homogeneous, like those of free-electron LLs. In contrast, LLs close to the VHS exhibit an inhomogeneous Berry curvature. $\Omega^{(z)}(\vec{k}) - \Omega_0$ shows extrema at the band extrema; for $E < 0$ (electron pockets), it is negative at the band maxima and positive at the minima. For $E > 0$ (hole pockets), this behavior is reversed.

Now we explain the large Chern number of the LL close to the VHS for odd $q$. The Berry curvature~\cite{gradhand2012first}
\begin{align*}
  \vec{\Omega}_n(\vec{k}) & = \mathrm{i} \sum_{m \ne n}\frac{\braket{u_n(\vec{k})|\nabla_{\vec{k}} H(\vec{k})|u_m(\vec{k})}\times(n\leftrightarrow m)}{[ E_n(\vec{k}) - E_m(\vec{k})]^2}
\end{align*}
of band $n$ is dominated by contributions from the adjacent bands. The maxima of band $n$ coincide with the minima of the adjacent band above, its minima coincide with the maxima of the adjacent band below. Viewing these avoided crossings as split Dirac points suggests to describe each avoided crossing by a two-band Hamiltonian 
\begin{align*}
  H & = \hbar\omega (-k_{x} \sigma_{x} + k_{y} \sigma_{y}) + m^{\star} \sigma_{z},
\end{align*}
in which $(k_{x}, k_{y})$ is taken relative to the $\vec{k}$ of the respective extremum. The nonzero effective mass $m^{\star}$ lifts the linear band crossing at $\vec{k} = 0$. Its sign determines the sign of the Berry curvature and, consequently, that of the Chern number $C = \pm \operatorname{sgn}(m^{\star}) / 2$ of the avoided crossing. Since the avoided crossings appear in even numbers, the (total) Chern number of the LL is integer.

As argued before, the sign of $m^{\star}$ corresponds to the fermion character. Therefore, the fermion character defines the sign of Berry curvature (minus $\Omega_0$). This argument fits to our numerical findings: below (above) the VHS, i.\,e., in the electron (hole) regime with $m^{\star} > 0$ ($m^{\star} < 0$), energy maxima coincide with minima (maxima) of the Berry curvature. As a result, the Berry curvature contributions of the maxima and minima of the LL oscillations cancel out and the Chern number of $-1$ is that of free-electron LLs.

The above reasoning does not hold for the LL near the VHS because its Berry curvature is dictated by states below \emph{and} above the VHS\@. Therefore, the dispersion minima are electronlike, which leads to a maximum of the Berry curvature. The maxima are holelike and, thus, also coincide with maxima of the Berry curvature (see panels~b and~c of Fig.~\ref{fig:VHS_square}). In total, the Berry curvature of this particular LL is positive throughout the BZ\@. Considering the two-band model for this LL, each of the $q$ minima and $q$ maxima induces a Chern number of $+1/2$. The total Chern number of the LL is thus $C = q$, from which  the Chern number of $-1$ due to the background $\Omega_0$ (free electrons) has to be subtracted. 

In summary, we obtain an outstanding Chern number of $C = q - 1$ for the LL at the VHS\@. A similar reasoning for even $q$ results in $C = q - 2$ for the LL pair close to the VHS\@. All other Landau levels carry the Chern number $-1$ of free-electron LLs because their Berry curvature is dictated by states with the same fermion character, with the consequence that minima and maxima contributions due to band oscillations cancel out (cf.\ the top and the bottom LL in Fig.~\ref{fig:VHS_square}c).

\subsubsection{An approximation for the Hall conductivity}
\label{sec:approximation}
The above line of argument lends itself to formulate an approximation for deriving Hall conductivities. This rule of thumb requires only knowledge of the zero-field band structure ($B = 0$).

\begin{figure}
  \centering
  \includegraphics[width=0.45\columnwidth]{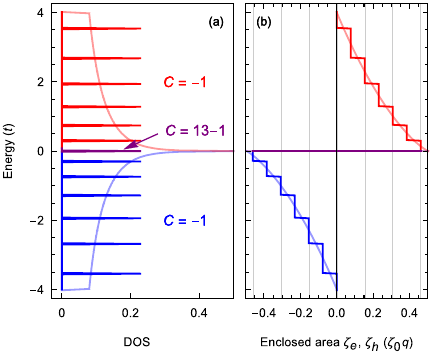}
  \caption{Quantum Hall effect on a square lattice with  $p/q = 1/13$. (a) The density of states (DOS, light smooth curve) is quantized into Landau levels (LLs; opaque peaks), each containing the same number of states. The LL at the VHS carries a Chern number of $C = q - 1 = +12$, the other ones $C = -1$. The fermion character is indicated by the color (cf.\ Fig.~\ref{fig:QHE_quad_cuts}). (b) The enclosed area $\zeta$ (proportional to the integrated DOS; light curves) is decomposed with respect to the fermion character ($\zeta_e$: electronlike, negative; $\zeta_h$: holelike, positive) and quantized (opaque). Energy in units of the hopping strength $t$.}
  \label{fig:QHE_quad_interpretation}
\end{figure}

If a Fermi line encloses an area $\zeta = (j+1/2) \, \zeta_0$ ($j$ integer, $\zeta_0=F/q$, $F$ area of the Brillouin zone) irrespective of the fermion character, a dispersionless LL is formed at the respective energy (Fig.~\ref{fig:QHE_quad_interpretation}a), according to Onsager's quantization scheme. All Landau levels carry Chern numbers of $-1$; an exception are the LLs close to the VHS which carry a large Chern number. Recall that at the VHS the Hall conductivity changes sign. This rough picture yields quite a detailed energy dependence of the Hall conductivity. Taking as an example a square lattice, the approximated conductivity (opaque in Fig.~\ref{fig:QHE_quad_interpretation}b) matches the numerically computed one (cf.\ Figs.~\ref{fig:QHE_quad_results}b and~\ref{fig:QHE_quad_interpretation}b). 

The semiclassical expression~\cite{lifshitz1957theory}
\begin{align}
  \sigma_{xy} & = \sigma_0 \, \left(\frac{\zeta_\mathrm{h}}{\zeta_0} - \frac{\zeta_\mathrm{e}}{\zeta_0}\right)
  \label{eq:lifshitz_sigma}
\end{align}
relates the Hall conductivity with the number of enclosed states and their fermion character (bright in Fig.~\ref{fig:QHE_quad_interpretation}b; $\zeta_\mathrm{e}$ for electrons, $\zeta_\mathrm{h}$ for holes); it reproduces the overall shape well but lacks quantization~\cite{arai2009quantum}. 

The above construction works well for free-electron LLs. However, it appears questionable for LLs close to the VHS because the Fermi lines are not closed (confer the Lifshitz transition at the constant-energy cut $\beta$ in Fig.~\ref{fig:QHE_quad_cuts}). Anyway, the rule describes the jump at the VHS if the fermion character is taken into account for the enclosed area $\zeta$ (taken negative for electrons and positive for holes). This corresponds to a shift of $+1$ at the VHS\@. Other lattice-induced features are not taken into account, for example the dispersion of the LLs. Still, the proposed approximation estimates well the overall shape of the Hall conductivity.

\subsubsection{Bulk-boundary correspondence} 
We now address the effect of the VHS on the topological edge states (TESs). The winding number of a band gap, Eq.~\eqref{eq:winding-number}, tells how many TESs bridge this gap.

\begin{figure*}
  \centering
  \includegraphics[width=0.8\textwidth]{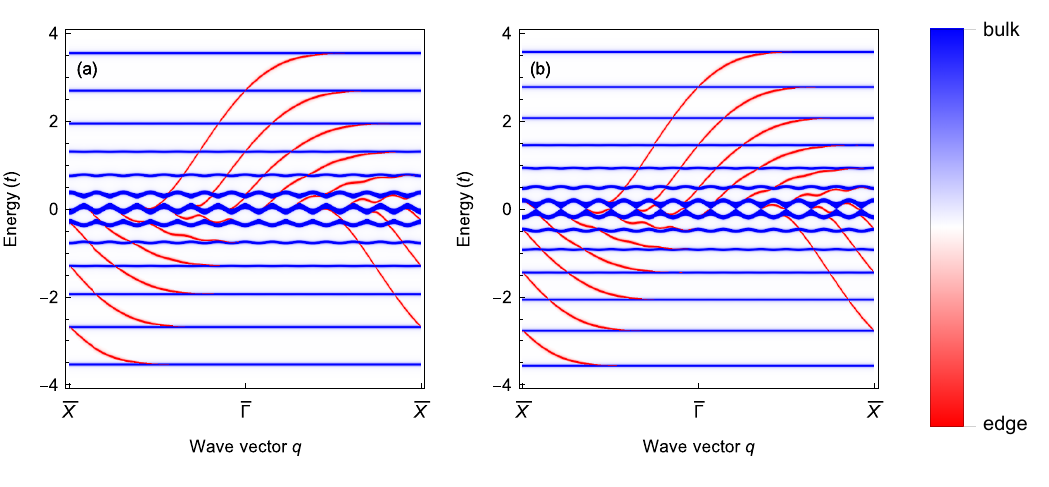}
  \caption{Landau levels and topological edge states of a square lattice with (a) $p/q = 1/13$ and (b) $p/q = 1/14$. The spectral density of a semi-infinite Hall bar is computed by Green function renormalization \cite{henk1993subroutine,bodicker1994interface}, the edge localization is coded by color (blue: bulk; red: edge). Energies in units of the hopping strength $t$.}
  \label{fig:edge_states}
\end{figure*}

Starting from the band bottom for odd $q$ ($q = 13$ in Fig.~\ref{fig:edge_states}a), the Chern number of $-1$ for each LL decreases the winding number by $1$. Consequently, the number of TES propagating to the left (with negative velocity) increases by $1$. At the VHS, the TESs cling to the oscillating LLs. The large Chern number in this region `compensates' all left-propagating TESs and creates $(q - 1) / 2$  right-propagating TESs (with positive velocity). Approaching the top of the band structure, the number of TESs decreases until it reaches zero.

The same holds for even $q$ ($q = 14$ in Fig.~\ref{fig:edge_states}b), with the exception that 
the winding number of the band gap at the VHS is zero. Hence, there are either no TESs at all or there are $q/2$ left- and $q/2$ right-propagating TESs. The latter is the case here: edge states from the bottom penetrate the lower band of the pair at $E_\mathrm{VHS}$ and edge states from above penetrate the upper band of the pair; then they cling to the other band. The Hall conductivity at $E_\mathrm{VHS}$ vanishes although there are edge states.

\subsubsection{Hierarchy of Landau levels}
\label{eq:quantization-p1}
In the definition of the flux $p / q = \Phi / \Phi_0$, $q$ defines the number of atoms in the magnetic unit cell and, therefore, fixes the number of LLs. While we focus on $p = 1$ in this Paper, a few remarks on the case $p>1$ will contribute to the discussion.

\begin{figure*}
  \centering
  \includegraphics[width=0.75\textwidth]{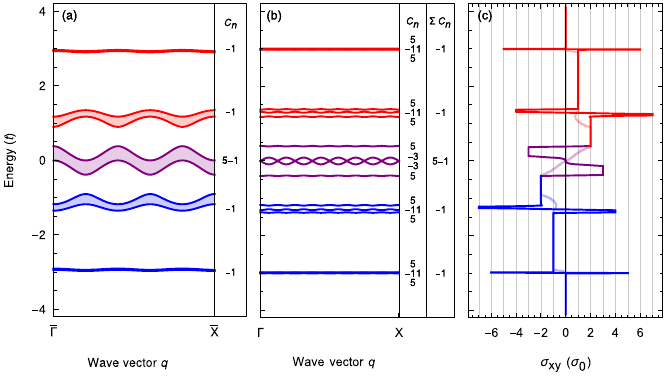}
  \caption{Quantum Hall effect on a square lattice. (a) Landau levels for $p / q = 1 / 5$. The fermion character is indicated by color (electronlike blue, holelike red). (b) Landau levels for $p / q = 3 / 16$. Chern numbers of the individual LLs are given in the column $C_{n}$, Chern numbers of the five groups in the column $\sum C_{n}$. (c) Hall conductivity $\sigma_{xy}$ for $p / q = 3 / 16$ (opaque lines) and $p / q = 1 / 5$ (light lines). Energies in units of the hopping strength $t$, $\sigma_{0} = e^{2} / h$.}
  \label{fig:QHE_quad_results_2}
\end{figure*}

For $p > 1$, one observes the formation of LL groups ($p / q = 3 / 16$ in Fig.~\ref{fig:QHE_quad_results_2}a), which is not described by Onsager's original quantization scheme. In an extended scheme $p / q$ is expressed as a continued fraction~\cite{chang1995berry,chang1996berry,hofstadter1976energy}
\begin{align*}
\frac{p}{q}=\frac{1}{f_1 + \frac{1}{f_2 + \frac{1}{f_3 + \cdots}}}, 
\end{align*} 
which establishes a hierarchy of LL groups. $f_1$ is the number of LL groups of order $1$, while the number of groups of higher order can be calculated from the $f_i$~\cite{chang1996berry}. With an unterminated  continued fraction even irrational values of $p/q$ can be calculated.

In our example
\begin{align*}
\frac{p}{q} & = \frac{3}{16} = \frac{1}{5+\frac{1}{3}},
\end{align*}
$f_{1} = 5$ tells that the LLs are arranged into $5$ groups of first order (Fig.~\ref{fig:QHE_quad_results_2}). These LL groups are related to those of the case $p / q = 1 / 5$ (reproduced in panel a). These bands split up in $\{3,3,4,3,3\}$ Landau levels (of second order). There are no LLs of higher order because the continued fraction is terminated.

The Chern numbers of the first-order LL groups concur with Onsager's quantization scheme~\cite{chang1996berry,avron1985quantization, thouless1982quantized}; the column $\sum C_{n}$ in Fig.~\ref{fig:QHE_quad_results_2}b can be produced from the approximation given in Section~\ref{sec:approximation}. These group Chern numbers are the sums of the Chern numbers of the individual LLs;  cf.\ the column $C_{n}$. Besides explicit calculation, the latter can be obtained from the Diophantine equation~\cite{dana1985quantised,thouless1982quantized,chang1996berry}. 

The hierarchy of LLs and their Chern numbers dictate the Hall conductivity; the first-order LL groups are dominating the overall behaviour [cf. opaque ($p/q=3/16$) and transparent ($p/q=1/5$) curves in Fig.~\ref{fig:QHE_quad_results_2}c].

\subsection{Quantum Hall effect on a triangular lattice}
\label{sec:QHEtriangular}
We turn briefly to the triangular lattice, extending the discussion of results given in  Ref.~\onlinecite{gobel2017THEskyrmion}. Due to the hexagonal symmetry of the triangular lattice the Hall conductivity is  unconventionally quantized, which is explained by the zero-field band structure.

In the gauge $\vec{A} = B (y-x/\sqrt{3}) \vec{e}_{x}$ the Hamiltonian reads
\begin{align*}
  H & =  t
  \begin{pmatrix}
  h_1& h_1^{(+)}	& 0& \dots	 & 0 &h_q^{(-)}      \\
  h_1^{(-)}	& h_2& h_2^{(+)}	& \dots  &0& 0 	  \\
  0&h_2^{(-)}	& h_3	& \dots &0 & 0 	  \\
  \vdots	& \vdots & \vdots 	& \ddots & \vdots& \vdots \\
  0 &0&0	& \dots & h_{q-1}&h_{q-1}^{(+)} \\
  h_q^{(+)} &0&0	& \dots &	h_{q-1}^{(-)} & h_q
  \end{pmatrix},
\end{align*}
with the matrix elements
\begin{align*}
  h_j & = 2 \cos\left( \frac{\sqrt{3}}{2}a k_x + \frac{1}{2} a k_y + 2 \pi \frac{p}{q} j \right),
\\
  h_j^{(+)} & = \left(h_{j}^{(-)}\right)^{\star} =\phi_{y}^{2} + \phi_{x} \phi_{y} \mathrm{e}^{- 2 \pi \mathrm{i} \frac{p}{q} \left( j + \frac{1}{2} \right)},
\end{align*}
and 
\begin{align*}
 \phi_{x} & \equiv \exp\left(\mathrm{i} \frac{\sqrt{3} a k_x}{2} \right),
 \\
 \phi_{y} & \equiv \exp\left(-\mathrm{i} \frac{a k_y}{2} \right).
\end{align*}
The lattice vectors are $\vec{a}_1 = a \vec{e}_{y}$ and $\vec{a}_2 = a (\sqrt{3} \vec{e}_{x} - \vec{e}_{y}) / 2$.

\begin{figure*}
  \centering
  \includegraphics[width=0.95\textwidth]{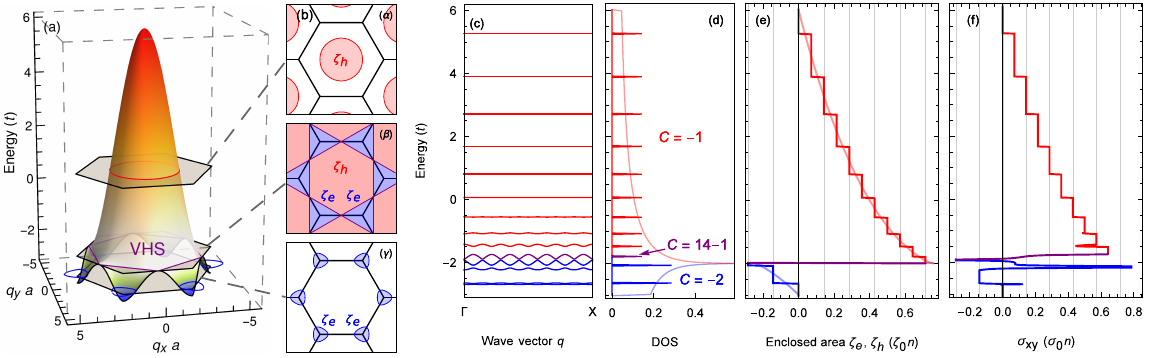}
  \caption{Quantum Hall effect on a triangular lattice for $p/q = 1/14$. (a) and (b) Zero-field band structure with constant energy cuts ($\alpha, \beta, \gamma$); as Fig.~\ref{fig:QHE_quad_cuts}. (c) Landau levels. The color represents the fermion character (blue electronlike, red holelike). (d) Density of states (DOS; semitransparent for the zero-field band structure shown in panel b) and Onsager-quantized levels (opaque; Chern numbers are indicated). The blue LLs appear in pairs and carry a total Chern number of $-2$. (e) Enclosed area in reciprocal space. The sign encodes the fermion character (electronlike negative, holelike positive). (f) Hall conductivity.  Energies in units of the hopping strength $t$, $\sigma_{0} = e^{2} / h$.}
  \label{fig:ueb_tria}
\end{figure*}

\begin{figure*}
  \centering
  \includegraphics[width=0.75\textwidth]{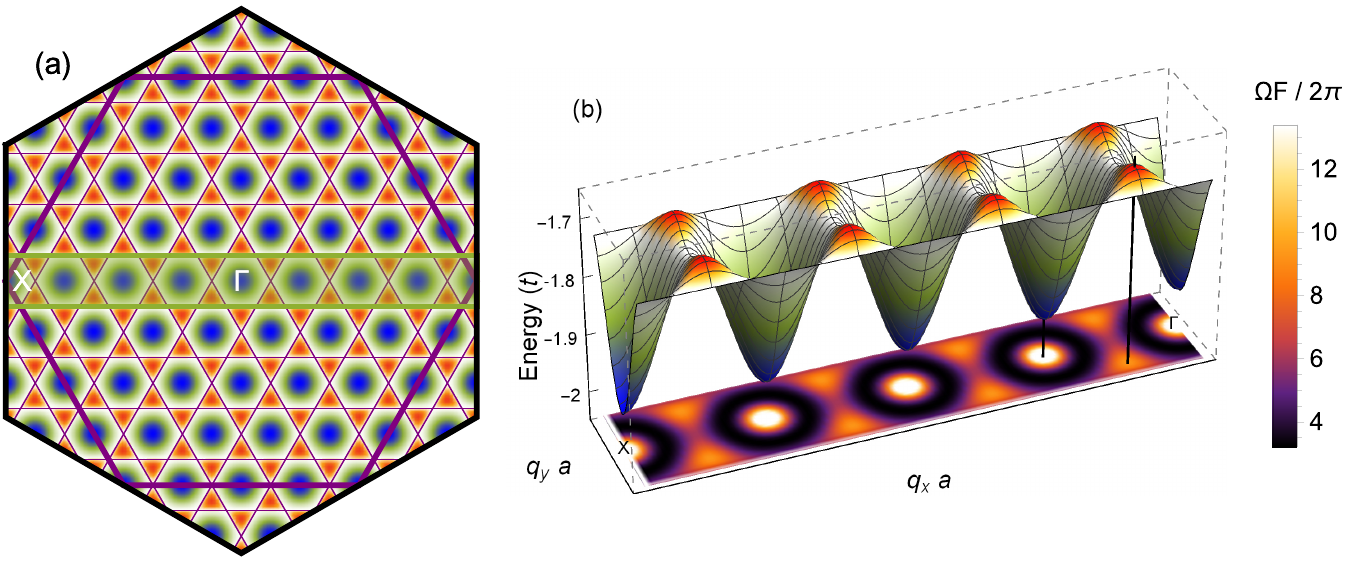}
  \caption{Landau level near the VHS of the triangular lattice for $q = 8$. (a) Oscillations of the Landau level represented in the structural Brillouin zone (BZ, hexagon). The magnetic Brillouin zone is depicted as green rectangle. (b) Threedimensional representation of the LL's dispersion in the magnetic BZ\@. Its Berry curvature $\Omega^{(z)} - \Omega_0$ is shown as color scale at the bottom (normalized to the area of the Brillouin zone $F$). Energies in units of the hopping strength $t$.}
  \label{fig:VHS_tria}
\end{figure*}

The quantum Hall effect is preferably described in a rectangular cell, similar to that of the square lattice; this facilitates a comparison of the two systems.  The magnetic Brillouin zone fits $q$-times into the structural Brillouin zone. The bands show $q$ oscillations that resemble the zero-field band structure (cf.\ Figs.~\ref{fig:ueb_tria}a and~\ref{fig:VHS_tria}) and appear most pronounced near the VHS (Fig.~\ref{fig:VHS_tria}c). There, the Chern number reads $q - 1$. For a detailed discussion we refer to Ref.~\onlinecite{gobel2017THEskyrmion}.

The zero-field band structure (Fig.~\ref{fig:ueb_tria}a) has two electron pockets at energies between the band bottom $E = -3 \, t$ and the VHS at $E_\mathrm{VHS} = -2 \, t$; otherwise it has one hole pocket (top of the bands at $E = +6 \, t$). Due to the different symmetry of the hexagonal lattice, the VHS is closer to the band bottom than to the top of the band; thus, a LL is not pinned exactly to the VHS\@. Nevertheless, the LL closest to the VHS is formed by states with positive \emph{and} states with negative effective mass, which causes the Berry curvature to be positive throughout the BZ (Fig.~\ref{fig:VHS_tria}b). This leads to the large Chern number of $C = q - 1$, as is the case for the square lattice.

The unconventional quantization of the conductivity shows up at energies below the VHS\@. There, each of the two separated electron pockets has to fulfill Onsager's quantization. Therefore, two LLs appear if $\zeta = 2\, (j + 1/2) \, \zeta_0$ ($j$ integer; Fig.~\ref{fig:ueb_tria}c and~d). The total Chern number of such a pair reads $-2$. As a result, the conductivity shows steps of $2 \, e^2 / h$ (Figs.~\ref{fig:ueb_tria}e and~f).

In summary, the interpretation of the QHE on the square lattice (Section~\ref{sec:QHEsquare}) can be carried over to the triangular lattice. A difference appears at energies below the VHS because the LLs are asymmetrically distributed about the VHS and two electron pockets (instead of one) show up.

\subsection{Topological Hall effect in skyrmion crystals}
\label{sec:THE}
The topological Hall effect (THE) in a skyrmion crystal is closely related to the QHE (Ref.~\onlinecite{gobel2017THEskyrmion}). A one-to-one correspondence has been established for bands and Chern numbers, except for energies close to VHSs. Nevertheless, even in this region the Hall conductivities of both effects are similar to each other in case of large skyrmions. 

The magnetic texture of a skyrmion (top hexagon in Fig.~\ref{fig:systems}c; notice that the spin in the center points in positive $z$ direction, what corresponds to a generating magnetic field in negative $z$ direction) carries an integer topological charge 
\begin{align*}
N_\mathrm{Sk}=\frac{1}{4\pi}\int_{xy} n_\mathrm{Sk}(\vec{r})\,\mathrm{d}^{2}r = w,
\end{align*} 
in which $w$ is the vorticity. It acts on the electron spin via a Zeeman interaction. The corresponding Zeeman term in the Hamiltonian can be transformed into a Peierls term whose effective magnetic field---the emergent field of the skyrmion (bottom hexagon in Fig.~\ref{fig:systems}c)---acts on the electron charge. This field is collinear with a nonzero average and, therefore, the Hall conductivity of the THE is similar to the conductivity of the QHE with a corresponding homogeneous magnetic field with $p/q = N_\mathrm{Sk}/n =\pm 1/n$, where $n$ is the number of atoms in the skyrmion unit cell.

Topological and quantum Hall effect differ in the inhomogeneity of the emergent field (THE: inhomogeneous; QHE: homogeneous). The inhomogeneity `bends' the almost flat LLs of the QHE and redistributes their Berry curvature. If it does not introduce level crossings, the Chern numbers remain unchanged. However near a VHS, the LLs are so close to each other that a bending \emph{could} alter the Chern numbers and, thus, the conductivity. Nevertheless, the total Chern number of a bundle of LLs  near the VHS is conserved. As a consequence, the conductivities of THE and QHE show the same global energy dependence.

Following Ref.~\onlinecite{hamamoto2015quantized}, the spin-dependent electronic structure is described by the tight-binding Hamiltonian
\begin{align} 
  H & = \sum_{ij} t \, c_{i}^\dagger \,c_{j} + m \sum_{i} \vec{s}_{i} \cdot (c_{i}^\dagger \vec{\sigma} c_{i})
  \label{eq:ham_the} 
\end{align}
($i$ and $j$ site indices) with real nearest-neighbor hopping strength $t$. The second sum couples the spins ($c_{i}^\dagger$ and $c_{i}$ are now two-component creation and annihilation operators) to the local magnetic texture $\{ \vec{s}_{i} \}$ of the skyrmion crystal via a Zeeman term; $\vec{\sigma}$ is the vector of Pauli matrices.

\begin{figure}
  \centering
  \includegraphics[width= 0.5\columnwidth]{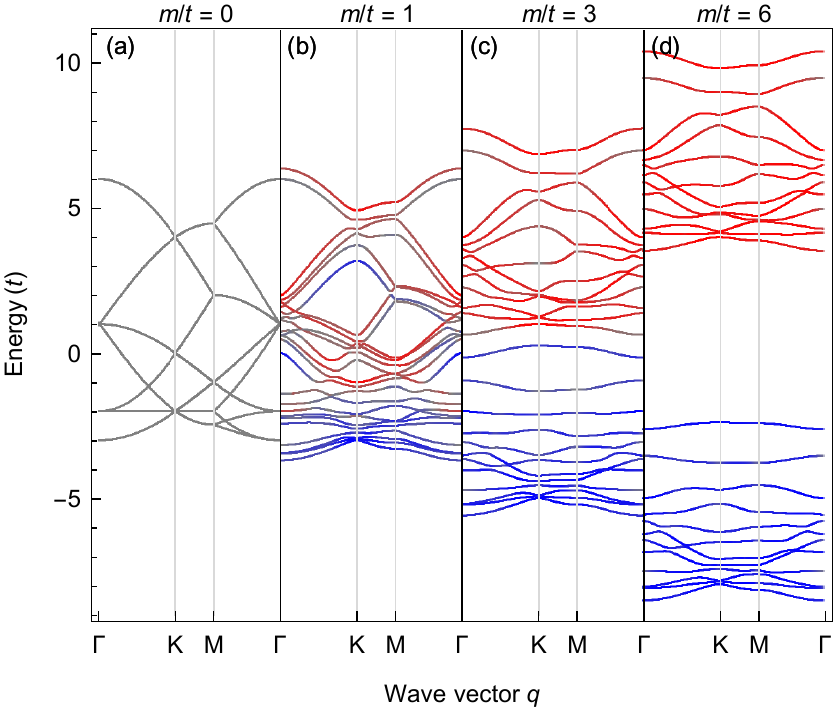}
  \caption{Band structure of a skyrmion crystal on a triangular lattice with $n = 12$ sites per magnetic unit cell for selected coupling strengths $m$ (in units of the hopping strength $t$; denoted at the top of each panel). For each band the alignment of the spin to the local magnetic texture is indicated by color (parallel: blue; antiparallel: red). Energies in units of $t$.}
\label{fig:pic_m_scan}
\end{figure}

The coupling strength $m$ to the magnetic texture is now discussed for a skyrmion on a triangular structural lattice with $n = 12$ atoms in the (magnetic) unit cell and $N_\mathrm{Sk}=1$. For $m = 0$, the spin is not coupled to the skyrmion texture and the bands are spin-degenerate. The band structure of the triangular lattice (Fig.~\ref{fig:ueb_tria}a) is back-folded into the magnetic Brillouin zone (Fig.~\ref{fig:pic_m_scan}a). The bands appear between $E = -3 \, t$ and $+6 \, t$, a van Hove singularity shows up at $E_\mathrm{VHS} = -2 \, t$.

For nonzero $m$, the spins align with the skyrmion texture and the spin degeneracy of the bands is lifted (Fig.~\ref{fig:pic_m_scan}b). Increasing $m$ further, the bands are separated into two blocks: one with spins parallel, the other with spins anti-parallel to the skyrmion texture (panels~c and~d). While for $m = 6 \, t$ both blocks deviate in details, in the limit $m \gg t$ both blocks exhibit identical, rigidly shifted dispersion relations. The band blocks, with width $\le 9 \, t$, are shifted by $\pm m$. While the upper bands of each block are well separated and quite dispersive, around the VHSs at energies $-2 \, t \pm m$ the band widths and gaps are considerably smaller. 

The band structures are invariant with respect to changes of the skyrmion helicity---when continuously turning a N\'{e}el- into a Bloch-type skyrmion---and the skyrmion number $N_\mathrm{Sk}$ (skyrmion and antiskyrmion).

For strong coupling ($m \gg t$), the electrons' spins are fully aligned with the skyrmion texture. Thus, it is sufficient to consider only one band block, which then describes spinless electrons; recall that this was also the case for the QHE discussed above. However, the effect of the skyrmion magnetic texture has to be taken into account by a local gauge transformation to the reference frame which is defined by the local magnetic moments \cite{everschor2014real, hamamoto2015quantized, ohgushi2000spin,gobel2017THEskyrmion}. The transformation is presented in Appendix~\ref{app:THEtransformation}.

In this emergent-field picture the THE is described by coupling of the electron's charge to a magnetic field in $\vec{z}$ direction. In the limit of large skyrmions, the magnetic texture is quasicontinuous and the emergent field can be understood as a real-space Berry curvature which is proportional to the skyrmion density $n_\mathrm{Sk}$ [eq.~\eqref{eq:skyrmiondensity}]. For small skyrmions, however, the discrete skyrmion density is proportional to the local ``spin chirality'', i.\,e., the solid angle spread out by neighboring spins. In contrast to the homogeneous magnetic field that causes a QHE, the emergent magnetic field for the THE is inhomogeneous (bottom hexagon in Fig.~\ref{fig:systems}c). Yet, it is nonzero on average, because it has to fulfill~\cite{everschor2014real}
\begin{align*}
 \frac{1}{4\pi\hbar} \int_\mathrm{uc} B^{(z)}(\vec{r}) \,\mathrm{d}^{2}r & = N_\mathrm{Sk},
\end{align*}
so that both effects (QHE and THE) can be compared. 

\begin{figure}
  \centering
  \includegraphics[width=0.5\columnwidth]{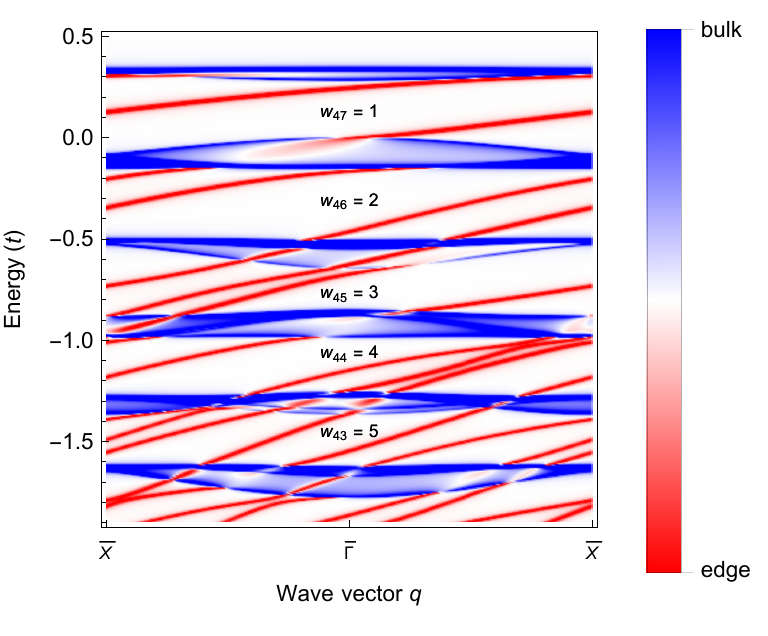}
  \caption{Bulk-boundary correspondence in a skyrmion crystal on a triangular structural lattice with $n = 48$ sites in the magnetic unit cell and $m/t=5$. The spectral density is depicted for bulk (blue) and edge (red) states in the energy range of well-separated bulk bands (band $43$ -- $48$). Each bulk state has a Chern number of $-1$, which reduces the winding number $w_{j}$ ($j = 43, \ldots, 48$) and, thus, the number of topological edge states in each band gap in steps of $1$.}
  \label{fig:edge_states_skyrm}
\end{figure}

The intimate relation of THE and QHE on a lattice becomes evident for the edge states. We briefly address edge states of a skyrmion lattice with $n = 48$ sites in the magnetic unit cell (Fig.~\ref{fig:edge_states_skyrm}), which corresponds to a QHE system with $p/q = 1/48$. The six topmost bands (blue) of the lower block are identified as LLs each of which carries a Chern number of $-1$. Hence, the number of edge states (red) bridging the band gaps increases by $1$ with decreasing energy. This explains the quantized topological Hall conductivity~\cite{hamamoto2015quantized,gobel2017THEskyrmion}.

\subsection{Approximation for the topological and quantum Hall conductivity}
\label{sec:approx}
The above interpretation of the QHE (Sections~\ref{sec:QHEsquare} and~\ref{sec:QHEtriangular}) suggests to formulate an approximation for the energy dependence of the transverse Hall conductivity, like in Refs.~\cite{lifshitz1957theory,arai2009quantum}. This rule applies also to the THE in a skyrmion crystal for strong coupling.

\subsubsection{Formulation of the approximation}
Consider a twodimensional lattice which either is subject to a homogeneous magnetic field or hosts a crystal of skyrmions with topological charge $N_\mathrm{Sk}$. Its unit cell comprises $n$ atoms. $n$ determines how the initial $b$ bands of the zero-field band structure are quantized. The zero-field band structure hosts $v$ van Hove singularities at energies $E_{\mathrm{VHS}}^{(i)}$, $i = 1,\ldots, v$ given by the symmetry of the structural lattice.

The approximation of the conductivity $\sigma_{xy}(E)$ proceeds as follows.
\begin{enumerate}
 \item The zero-field band(s) is (are) quantized in accordance with Onsager's quantization prescription. The resulting Landau levels are assumed dispersionless. Their homogeneously distributed Berry curvature yields a Chern number of $- N_\mathrm{Sk}$ ($+N_\mathrm{Sk}$) per band in the lower (upper) band block for the THE in a skyrmion crystal. For the QHE with a uniform field one sets $N_\mathrm{Sk} = \sign(B)$ and treats only the lower block.
 \item The conductivity is shifted at the VHSs $E_{\mathrm{VHS}}^{(i)}$ by $\pm N_\mathrm{Sk} \sigma_0 n  / v$ for the lower and upper block, respectively, to account for the lattice influence ($\sigma_0=e^2/h$). For the QHE the shift is $\sign(B)\sigma_0 n  / v$.
\end{enumerate}
This procedure yields the conductivities
\begin{align} \label{eq:deduction}
  \begin{split}
  \sigma_{xy}(E) &
   = 
  \mp \frac{e^{2}}{h} N_\mathrm{Sk}  \left\lfloor n \int_{-\infty}^{E} D(E') - \frac{1}{v} \sum_{i}^{v} \theta(E' -E_\mathrm{VHS}^{(i)}) \,\mathrm{d}E' \right\rfloor_{o(E)} 
  \end{split}
\end{align}
for the lower ($-$) and the upper ($+$) band block. $D(E)$ is the normalized density of states, $\int D(E') \,\mathrm{d}E = b$, and $\theta$ is the Heaviside function. $o(E)$ counts the number of pockets at $E$ and redefines the conventional floor function: $\lfloor x \rfloor_o$ rounds down in steps of $o$, while the conventional $\lfloor x \rfloor$ gives the next lower integer of $x$. Thus, $\lfloor x \rfloor_{o(E)}$ accounts for Onsager's quantization scheme, in which the winding number at the next lower $E_\mathrm{VHS}$ determines the offset of the integer quotient. An offset of $o/2$ has to be included to account for the $1/2$ in the Onsager scheme [LL formation if $\zeta = (j + 1/2) \,\zeta_0$] per electron- or hole-pocket.

\paragraph*{Addition.} For a skyrmion crystal, the band energies have to be scaled by $\cos ( \overline{\vartheta_{ij}} / 2)$ to adjust the total band width [see also eq.~\eqref{eq:teff} in Appendix~\ref{app:THEtransformation}]. The average angle of neighboring spins can be approximated by $\overline{\vartheta_{ij}} = 3 /  2 \cdot \pi a / \lambda$. Here, $\lambda$ is the pitch of the spin spirals whose superposition forms the skyrmion crystal (cf.\ Ref.~\onlinecite{okubo2012multiple}). For large skyrmions ($\lambda \gg a$) the scaling factor approaches $1$ and the scaling is irrelevant.

\subsubsection{Application to the honeycomb lattice} 
\begin{figure*}
  \centering
  \includegraphics[width=0.95\textwidth]{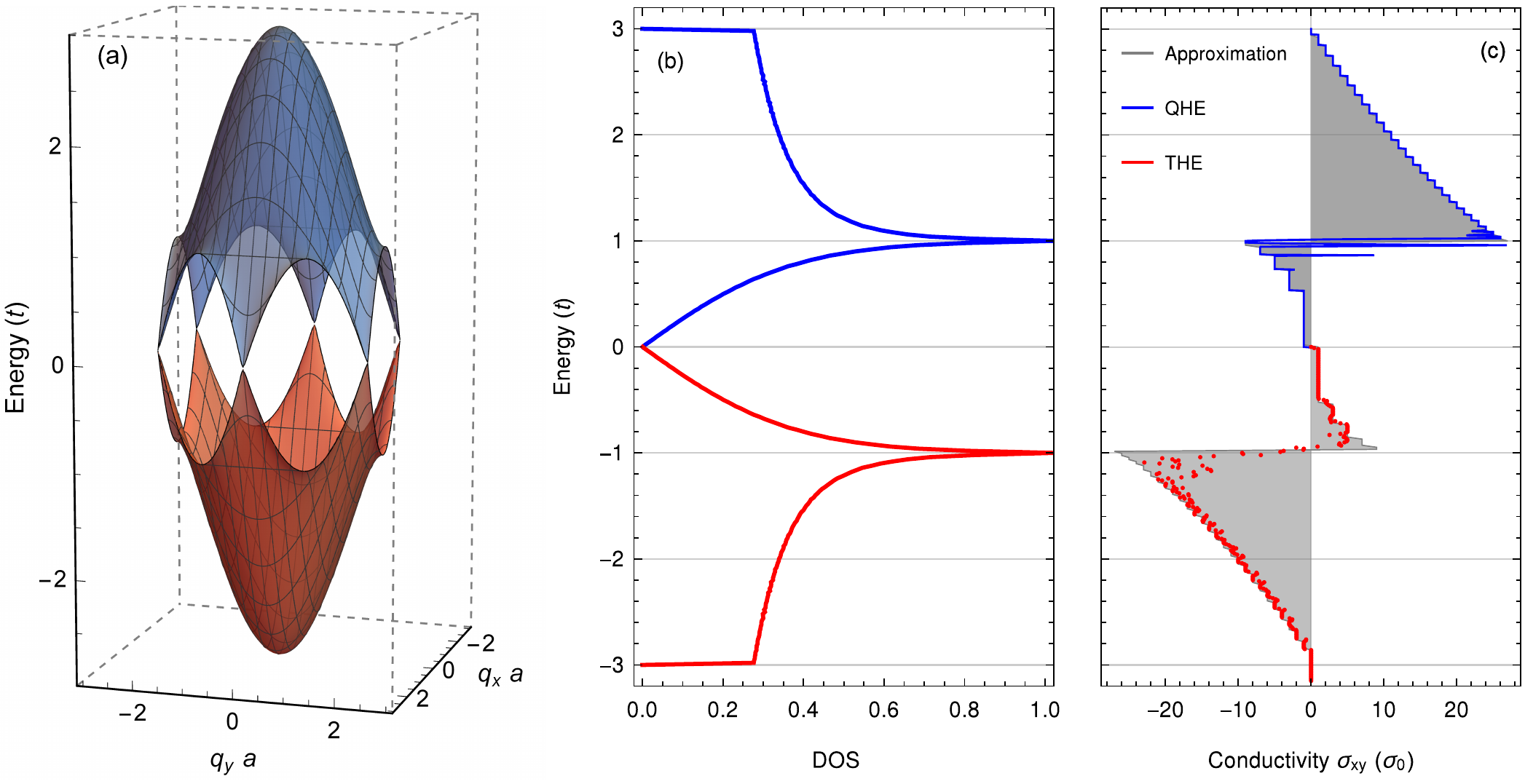}
  \caption{(Color online) Deduction of the topological Hall conductivity and quantum Hall conductivity from (a) band structure and (b) density of states of a honeycomb lattice for zero magnetic field. (c) Lower block of the topological Hall conductivity $\sigma_{xy}$ \textit{versus} Fermi energy for a skyrmion crystal on a honeycomb lattice (red) and QHE with uniform magnetic field (blue) for $n = 72$ sites in the magnetic unit cell. For clarity, only the lower or upper halves are depicted. The topological Hall conductivity was computed for very strong coupling ($m = 900\, t$) and rigidly shifted by $m$. The approximation eq.~\eqref{eq:deduction} produces the gray curve.  Energies in units of the hopping strength $t$, $\sigma_{0} = e^{2} / h$.}
  \label{fig:honeycomb_band_dos}
\end{figure*}

As an illustration, we apply eq.~\eqref{eq:deduction} to the honeycomb lattice. The two atoms in the structural unit cell yield $b = 2$ bands and a DOS that is symmetric about the VHS (panels~a and~b of Fig.~\ref{fig:honeycomb_band_dos}). Considering $q$ sites in each of the sublattices yields $n = 2 \, q$. The QHE Hamiltonian then reads in matrix form
\begin{align*}
  H = t\begin{pmatrix}
0& h_1^{(+)}	& 0& 0&\dots	 &0& 0 &g^{(-)}      \\
h_1^{(-)}	& 0& g^{(+)}	&0& \dots &0  &0& 0 	  \\
0&g^{(-)}	& 0	&h_2^{(+)}& \dots &0&0 & 0 	  \\
0&0&h_2^{(-)}	& 0& \dots &0&0 & 0 	  \\
\vdots	& \vdots & \vdots & \vdots& \ddots 	& \vdots & \vdots& \vdots \\
0 &0&0	& 0&\dots &0&g^{(+)}&0 \\
0 &0&0	& 0&\dots &g^{(-)}& 0&h_q^{(+)} \\
g^{(+)} &0&0	& 0&\dots &0 &	h_q^{(-)} & 0
\end{pmatrix},
\end{align*}
with
\begin{align*}
  h_j^{(+)} & = \left(h_j^{(-)}\right)^{\star} =  \phi_{x} \phi_{y} + \left(\phi_{x}^2\right)^{\star} \mathrm{e}^{- 2 \pi \mathrm{i} j \frac{p}{q}},\\
  g^{(+)} & = \left(g^{(-)}\right)^{\star} = \phi_{x}^{\star} \phi_{y},
\end{align*}
and 
\begin{align*}
 \phi_{x} & \equiv \exp\left(\mathrm{i} \frac{a k_x}{2\sqrt{3}} \right),
 \\
 \phi_{y} & \equiv \exp\left(\mathrm{i} \frac{a k_y}{2} \right).
\end{align*}

To check the validity of the approximation we compare the Hall conductivities of the THE and of the QHE---computed from the above Hamiltonian---with that produced by eq.~\eqref{eq:deduction} for $n = 72$  and for large coupling $m$. Since in all cases $\sigma_{xy}(E)$ is antisymmetric to the block center we show the THE data only for the lower half and QHE data only for the upper half of the band block.

All three data sets agree well, even the quantization plateaus are reproduced in a wide energy range (Fig.~\ref{fig:honeycomb_band_dos}c). The quantum Hall conductivity (blue) matches the approximation (gray). Deviations show up close to the VHS at which sizable modulations indicate dispersive LLs; recall that the LL dispersion is not taken into account by the approximation. The jump itself is reproduced best for large $n$. For small $n$ the jump may be shifted in energy because it is not pinned to the VHS (cf.\ the triangular lattice in  Section~\ref{sec:QHEtriangular}). In the approximation, however, the jump is introduced artificially at $E_\mathrm{VHS}$, what explains the deviation.

Larger deviations appear for the THE (red). Again, the plateaus are reproduced by the approximation but not in great detail. The inhomogeneous emergent field of the skyrmion texture creates band bendings even for the former dispersionless LLs of the QHE far off the VHSs. Therefore, the conductivity shows small modulations at the plateau edges. In addition, lattice effects are significant at $E_\mathrm{VHS}$, as for the QHE; the jump is quite broad, spread over several levels.

The THE data have been scaled by $\cos ( \overline{\vartheta_{ij}} / 2) = \cos(\pi/12)$ [eq.~\eqref{eq:teff} in Appendix~\ref{app:THEtransformation}], as explained in the previous Section (see `\textit{Addition}'). This scaling is insignificant for skyrmions of size $n = 72$, as can be seen by the minimal shift of $E_\mathrm{VHS}$ (compare the sign change of the gray curve in the lower and upper halves of the block in Fig.~\ref{fig:honeycomb_band_dos}c).

Summarizing, the approximation works well at energies apart of VHSs: both the quantization steps and the change of sign are reproduced for QHE and THE\@. Close to VHSs the approximation of the THE improves with skyrmion size. 

\section{Conclusion and outlook}
\label{sec:conclusion}
In this Paper we discussed the quantum and the topological Hall effect (QHE, THE) on square, triangular, and honeycomb lattices, with a focus on the energy dependence of the quantized Hall conductivities $\sigma_{xy}$. A sizable jump in $\sigma_{xy}$, which is accompanied by a change of sign, is attributed to van Hove singularities of the zero-field band structure, as for the QHE in Refs.~\onlinecite{hatsugai2006topological,arai2009quantum}. We showed that this sign change can be traced back to a single band with a very large Chern number. While the bands below and above a van Hove singularity have positive as well as negative Berry curvature contributions, they exclusively induce positive Berry curvature in the distinguished band in the vicinity of the singularity.

We pointed out that the THE is closely related to the QHE, because the emergent magnetic field due to the skyrmion texture is nonzero on average. Thus, the topological Hall conductivity shows an energy dependence similar to that of the quantum Hall conductivity. To support our results we calculated Chern numbers and winding numbers as well as the topological edge states.

Based on our findings we developed a handy approximation for the Hall conductivity. This approximation is in good agreement with the quantum Hall and topological Hall results. For the quantum Hall effect, it only lacks effects of band oscillations near the van Hove singularity. Our approximation gives non-specialists a rule of thumb to determine the transverse Hall conductivity of both effects for any structural lattice. It circumvents computationally demanding calculations of the Berry curvature.

Concerning experiments, the prominent features of $\sigma_{xy}$---quantization, jump, and change of sign---are preferably investigated in the THE of skyrmion crystals. Skyrmion crystals with a skyrmion radius of about $\unit[1]{nm}$ [e.\,g., Fe/Ir(100), Ref.~\onlinecite{heinze2011spontaneous}] would act as  gigantic emergent magnetic fields of about $\unit[4000]{T}$~\cite{hamamoto2015quantized}. This way, the predicted properties of the QHE in lattices could be (indirectly) reviewed because of the relation of THE and QHE: results for the THE can be carried over to the QHE and \textit{vice versa}. 

The THE can be studied in metals which host a skyrmion crystal (SkX) phase, e.\,g., MnSi (Ref.~\onlinecite{muhlbauer2009skyrmion}), Fe$_{1-x}$Co$_{x}$Si (Ref.~\onlinecite{yu2010real}), and FeGe (Ref.~\onlinecite{yu2011near}). Typically, finite temperatures and external magnetic fields $B$ are necessary to stabilize  a SkX phase. In samples with negligible anomalous Hall effect---another additional contribution to the Hall conductivity, that is significant for sizable intrinsic spin-orbit coupling---the Hall conductivity $\sigma_{xy}$ increases linearly with $B$ if $B$ is small. A transition from a topologically trivial phase to a SkX phase would cause a sharp increase of $\sigma_{xy}$ because the THE sets in abruptly at the phase boundary~\cite{gobel2017THEskyrmion}. This additional contribution depends on the lattice geometry, as shown in this Paper. It changes sign when the chemical potential passes a van Hove singularity, which could be achieved by applying a gate voltage. At low temperatures, the THE signal exhibits its salient features (quantization and change of sign) most clearly. Note that gating is limited to several hundred $\text{meV}$ in experiments. Thus, to experimentally verify the sign change, the ungated chemical potential should lie in the vicinity of a van Hove singularity. This restriction is lifted for the detection of the quantization steps which occur in the entire energy range.

Depending on the desired feature, different sizes of skyrmions are favorable in experiments: the change of sign in $\sigma_{xy}$ shows up sharply for large skyrmions, as the integrated density of states is well resembled by the numerous Landau levels~\cite{gobel2017THEskyrmion}. A compromise between sharpness of the sawtooth-shaped feature (favored by large skyrmions) and signal strength (favored by small skyrmions) has to be made, because a conductance rather than a conductivity is measured in an experiment. Low-temperature skyrmion crystals with very small skyrmions are preferable to detect the unconventional quantization. The quantization plateaus of $\sigma_{xy}$ are largest for small skyrmions~\cite{gobel2017THEskyrmion} and low temperatures prevent smoothening of the corresponding steps. The preparation of such SkXs is challenging but feasible \cite{von2014interface,heinze2011spontaneous,brede2014long,hagemeister2016skyrmions,wiesendanger2016nanoscale}.

A combined analysis of the anomalous and the topological Hall effects seems to be worthwhile in the future. If  intrinsic spin-orbit coupling is sizable, the anomalous contribution to the Hall effect has to be calculated to extract the topological contribution from experimental data~\cite{matsuno2016interface,schulz2012emergent,porter2014scattering}.

The skyrmion texture affects magnon transport as well. The transformation that produces the emergent field for electrons leads to an emergent electrodynamics that transforms the Landau-Lifshitz-Gilbert equation into the Hamiltonian of a charged particle in fictitious fields~\cite{gungordu2016theory}. This suggests to apply  the argumentation of this Paper to analogs for the topological magnon Hall effect~\cite{van2013magnetic,mochizuki2014thermally, mook2017magnon}. 

Our approximation suggests an expansion to skyrmion crystals with $|N_\mathrm{Sk}|>1$, since the ratio $p/q$ that defines the magnetic field of the QHE is related to $N_\mathrm{Sk}/n$ for the THE. Single skyrmions~\cite{leonov2015multiply}, as well as crystals with a higher skyrmion number~\cite{ozawa2017zero} have already been simulated. The recently predicted antiferromagnetic skyrmions on square lattices \cite{zhang2016antiferromagnetic,PhysRevLett.116.147203,PhysRevB.95.054421} lend themselves for studying the topological spin Hall effect in detail or to find a way to generate a nonzero THE by making the two skyrmion sublattices inequivalent.

\begin{acknowledgments}
This work is supported by Priority Program SPP 1666 of Deutsche Forschungsgemeinschaft (DFG).
\end{acknowledgments}

\appendix

\section{Mapping the topological onto the quantum Hall effect}
\label{app:THEtransformation}
The local gauge transformation addressed in Section~\ref{sec:THE} is mediated by the gauge field $\vec{A}(\vec{r})$ that defines the emergent magnetic field $\vec{B}(\vec{r}) = \vec{\nabla} \times \vec{A}(\vec{r})$. $\vec{B}(\vec{r})$ is along the $z$ direction and inhomogeneous. Its $z$ component \cite{everschor2014real}
\begin{align}
  B^{(z)}(\vec{r}) & = \frac{1}{2} \vec{s}(\vec{r}) \cdot \left(\frac{\mathrm{d} \vec{s}(\vec{r})}{\mathrm{d} x} \times \frac{\mathrm{d} \vec{s}(\vec{r})}{\mathrm{d} y}\right)
  \label{eq:emergent-field}
\end{align}
($\hbar = 1$) is given by the topological charge density of a skyrmion (lower hexagon in Fig.~\ref{fig:systems}c). Since the emergent field couples to the charge but not to the spin of the electron, the gauge transformation recasts the coupling of an electron's spin to the magnetic texture as a fictitious field acting on its charge. Both descriptions are equivalent and yield identical results for the THE, in the limit $m\to \infty$\@.

In a tight-binding model, the gauge field introduces new effective hopping strengths
\begin{align*}
  t^\mathrm{eff}_{ij} & = \tilde{t}_{ij} \,\exp\left( -\mathrm{i} e / \hbar \int_{\vec{r}_{i} \to \vec{r}_{j}} \vec{A}(\vec{r}) \cdot \mathrm{d}\vec{l} \right).
\end{align*}
$\mathrm{d}\vec{l}$ points along the hopping path from site $i$ to site $j$. $t^\mathrm{eff}_{ij}$ is the hopping strength in the QHE Hamiltonian
\begin{align*}
	H_\mathrm{QHE} & = \sum_{ij} t^\mathrm{eff}_{ij} \, d_{i}^\dagger \, d_{i}
\end{align*}
for the QHE, in which $d_{i}^\dagger$ ($d_{i}$) is a creation (annihilation) operator. The effective hopping can be expressed as~\cite{hamamoto2015quantized}
\begin{align}
  t^\mathrm{eff}_{ij} & = t \cos{\frac{\vartheta_{ij}}{2}} \mathrm{e}^{\mathrm{i} a_{ij}},
 \label{eq:teff}
\end{align}
with $t$ from eq.~\eqref{eq:ham_the} and $\vartheta_{ij}$ the angle between the spins of site $i$ and site $j$. With the corresponding polar angles $\varphi_{i}$ and $\varphi_{j}$, the phase reads
\begin{align}
  a_{ij} & = \arctan\frac{-\sin(\varphi_{i} - \varphi_{j})}{\cos(\varphi_{i} - \varphi_{j})+\cot\frac{\vartheta_{i}}{2} \cot\frac{\vartheta_{j}}{2}}. \label{eq:discreteaij}
\end{align}

\bibliography{short,MyLibrary}
\bibliographystyle{apsrev}
\end{document}